 \documentclass{aa}  
 \usepackage{graphicx}
 \usepackage{txfonts}
\begin{document}

 \newcommand{\vdag}{(v)^\dagger}
 \newcommand{\oneskip}{\vskip\baselineskip}
 \newcommand{\cc}{\hbox{$\mu$}}
 \newcommand{\ergsec}{\hbox{erg s$^{-1}$}}
 \newcommand{\aaa}{\hbox{M$_{\odot}$}}
 \newcommand{\bb}{\hbox{$\nu$}}
 \newcommand{\dd}{CXOU J205847.5+414637}
 \newcommand{\ee}{GRO J2058+42 }
 \newcommand{\dy}{\hbox{d$^{-1}$}}

 \def\hr{\hbox{$^{\rm h}$}}                 
 \def\fhr{\hbox{$.\!\!^{\rm h}$}}           
 \def\deg{\hbox{$^\circ$}}                  
 \def\fdeg{\hbox{$.\!\!^\circ$}}            
 \def\sec{\hbox{$^{\rm s}$}}                
 \def\fsec{\hbox{$.\!\!^{\rm s}$}}          
 \def\arcm{\hbox{$^\prime$}}                
 \def\farcm{\hbox{$.\mkern-4mu^\prime$}}    
 \def\arcs{\hbox{$^{\prime\prime}$}}        
 \def\farcs{\hbox{$.\!\!^{\prime\prime}$}}  
 \def\fday{\hbox{$.\!\!^{\rm d}$}}          
 \def\per{\hbox{$^{\rm p}$}}                
 \def\fper{\hbox{$.\!\!^{\rm p}$}}          
 \def\mm{\hbox{$^{\rm m}$}}                 
 \def\fmm{\hbox{$.\!\!^{\rm m}$}}           


 \title{Optical variabilities in Be/X-ray binary system:}
 \subtitle{\ee (\dd)}
 \author{\"U. K{\i}z{\i}lo\u{g}lu,
         N. K{\i}z{\i}lo\u{g}lu,
         A. Baykal, 
         S.K. Yerli, 
         M. \"Ozbey  
        } 


  \institute{Physics Department, Middle East Technical University, Ankara 06531, Turkey
 }

 \date{}

\abstract
 {}
{We present an analysis of long-term optical monitoring observations and
optical spectroscopic observations of the 
counterpart to \dd ~(high mass X-ray binary system).
We search for a variability in the light curve of Be star.
    } 
{We used differential magnitudes in the time series analysis. 
The variability search in the optical light curve was
 made by using different algorithms.
The reduction and analysis of spectra were done by using MIDAS and
 its suitable packages.
    }
{
We have performed a frequency search which gave us the
value 2.404  \dy. This value is attributed to the non-radial pulsation
of Be star. H$\alpha$ emission line profiles
always show double-peaked emissions with a mean equivalent width of
2.31$\pm$0.19 \AA ~and a peak separation of 516$\pm$45 km/s.
This suggests that Be star disk is still present. \dd ~is in
X-ray quiescent state.
    }
 {}


 \maketitle

\section{Introduction}
Be/X-ray binaries consist of a Be star and a neutron star
 which has little influence
on the Be star. Be stars are on or just off the Main Sequence and they
have rapid rotational velocities. Majority
 of them are found to rotate at  0.7 of their break-up velocity
 (Porter $\&$ Rivinius 2003).
H$\alpha$ emission line and infrared excess are the observational
 characteristics of Be stars. They show disappearance and re-appearance
of emission lines. Be stars often display enhancement or fading of their
brightness. It is thought that enhancements in brightness are associated
with the mass loss episodes and appear to be induced by non radial
pulsations (Rivinius et al.\ 2003).
 Studies of spectroscopic variations in some line profiles
 are interpreted as due to non radial pulsations (Rivinius et al.\ 2003,
 Neiner et al.\ 2005).
A Be star has a dense disk in its equatorial
 plane (Quirrenbach 1997; Waters 1986) which
 is fed from the material expelled
from the fast rotating Be star due to radiatively driven wind or photospheric
pulsations (Porter $\&$ Rivinius 2003). Among the disk models the favorable
one is the viscous disk model (Lee, Osaki $\&$ Saio 1991; Okazaki 2001).
Quasi Keplerian disks are held by viscosity (Okazaki $\&$ Negueruela 2001;
Okazaki 2001). Angular momentum is transferred from the inner regions
of the disk towards the outer region by the viscosity.
This disk causes  the X-ray outburst of NS either at the
periastron  passage (Type I outburst)
 or alters the outer part of disk (Type II outburst) without showing
 any correlation with orbital parameters (Negueruela  2004).
X-rays are produced as a result of accretion of matter onto the NS.
 In a Be/X-ray system there can be X-ray quiescence periods although a Be
disk is present (Negueruela $\&$ Okazaki 2001; Negueruela et al.\ 2001).

GRO J2058+42 (\dd), a transient 198 s X-ray pulsar was discovered by the BATSE
 instrument on the Compton Gamma Ray Observatory during a giant outburst
in September 1995 (Wilson et al.\ 1996).
Wilson et al.\ (1998) proposed that GRO J2058+42 was undergoing
 periastron and apastron outbursts in a 110 day orbit.
 They have estimated the/its? luminosity as  0.2-1X$10^{37}$ erg/s and
 no optical counterpart has been identified.
Corbet et al.\ (1997) showed the presence of modulation with a
period of approximately 54 days in the analysis of RXTE/ASM (Rossi X-ray Timing
Explorer/All Sky Monitor)
data obtained between 1996 and 1997.
Reig et al.\ (2004a) performed optical photometric and spectroscopic
observations of the best fit to GRO position of the GRO J2058+42. They
suggested that the star was located at $\alpha$= 20$^{h}$ 58$^{m}$ 47$^{s}$,
  $\delta$= +41$^{\circ}$ 46$'$ 36$''$,
 as the most likely optical counterpart. Spectra of the proposed
 star showed a double peak H$\alpha$ emission
profile with a mean equivalent width of 4.5 \AA
 ~which supports its classification as Be/X-ray binary.

Subsequent Chandra observations did not find a source within the GRO
positional error-box, but just outside, named \dd.
Wilson et al.\ (2005) obtained optical observations of CXOU J205847.5+414637.
Its optical spectrum contained a strong H$\alpha$ line in
 double peaked emission. They proposed that CXOU J205847.5+414637 and
GRO J2058+42 are the same object. They classified the spectral type
of this object as O9.5-B0IV-V
(V=14.9).
They estimated the distance as 9$\pm$1.3 kpc.
When RXTE/ASM data for GRO J2058+42 were folded at 55.03 d period,
pulsation from this source were detectable for 1996-2002 data,
but were not present in 2003-2004 observations.
 This system was in X-ray quiescent phase since 2002.
During the X-ray quiescence phase
the Be disk was present since H$\alpha$  was in emission as seen
from their  spectroscopic study (Wilson et al.\ 2005).
 \begin{figure*}
 \centering
  \includegraphics[clip=true,scale=0.5,angle=270]{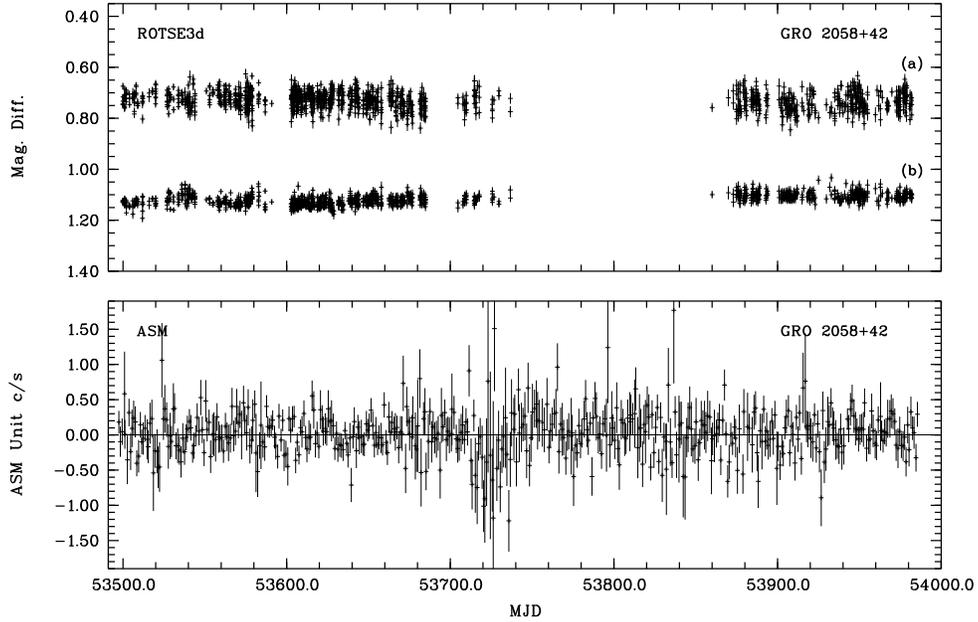}
 \caption{ROTSEIIId differential light curve of Be/X-ray binary system
          (top panel (a)) and mean light curve of
          reference stars properly offsetted (top panel (b))  ~for
          the period  2005-2006.
              X-ray light curve of
             the system obtained from RXTE/ASM data is given in
             the bottom panel for the same time interval as
             ROTSEIIId observations.} 
 \label{fig1}
 \end{figure*}
 \begin{table}
 \centering
 \caption{\dd ~and photometric reference stars used.}
 \label{table:1}
 \begin{tabular}{@{}ccccc@{}}
 \hline\hline
  Star  & R.A.      & Decl.     &  USNO A2  \\
        & (J2000.0) & (J2000.0) &  R mag    \\ 
 \hline
 J205847.5+414637  & 20\hr 58\mm 47\fsec54 & +41\deg 46\arcm 37\farcs3 & 14.4  \\
 Star 1 & 20\hr 58\mm 53\fsec53 & +41\deg 46\arcm 28\farcs0 & 13.9  \\
 Star 2 & 20\hr 58\mm 45\fsec85 & +41\deg 45\arcm 06\farcs0 & 13.9  \\
 Star 3 & 20\hr 59\mm 05\fsec50 & +41\deg 44\arcm 20\farcs1 & 14.1  \\
 \hline
 \end{tabular}
 \end{table}

There is no long term optical
monitoring for this system, therefore in this study
we present optical observations obtained by ROTSEIIId
 telescope. 
The coordinates of \dd ~were taken from Reig et al.\ (2004a) and
Wilson et al.\ (2005). 
We present optical spectroscopic observations obtained
in 2006. We report short term variabilities seen in the optical light
 curve. From the analysis of archival RXTE/ASM observations we attempt to detect
 the orbital modulation in 2005 and 2006.

\section{Observations and Data reduction}

The optical data were obtained with Robotic Optical Transient Experiment
\footnote{http://www.rotse.net}
(ROTSEIIId)  and  Russian-Turkish 
1.5 m Telescope \footnote{http://www.tug.tubitak.gov.tr} (RTT150)
located at Bak{\i}rl{\i}tepe, Antalya, Turkey.

\subsection{Optical Photometric observations}

The CCD observations  were obtained  between Jan, 2005 and
Aug, 2006 with 45 cm ROTSEIIId robotic reflecting telescope.
ROTSEIII telescopes which operate without filters were described
 in detail by Akerlof et al.\ (2003).
ROTSEIIId has equipped with a  2048$\times$2048 pixel CCD.
 The pixel scale is 3.3 arcsec
per pixel for a total field of view
 1.$^{\circ}$85$\times$1.$^{\circ}$85. A total of about 1440 CCD
 frames were collected during the observations.
In 2006, we have $\sim$400 data points.
 Due to the other scheduled observations and atmospheric conditions
 we have obtained 1--20 frames at each night
with an exposure time of 20 sec.
 All images were automatically dark- and flat-field corrected
by a pipeline as soon as they were exposed. Dark field frames which were 
accumulated each night were used in the data reduction pipeline together with 
proper sky-flat and fringe frames.  
 For each corrected image, aperture photometry was applied using 5 pixel
(17 arcsec) diameter aperture to obtain the instrumental magnitudes.
 These magnitudes were calibrated by comparing all the field stars against
 USNO A2.0 R-band catalog.
 Barycentric corrections were made to the times of
each observation by using JPL DE200 ephemerides prior to the analysis with the
period determination methods.
Details on the reduction of data were described in K{\i}z{\i}lo\u{g}lu,
K{\i}z{\i}lo\u{g}lu $\&$ Baykal (2005)
and Baykal, K{\i}z{\i}lo\u{g}lu $\&$ K{\i}z{\i}lo\u{g}lu (2005).

We used differential magnitudes in the time series analysis.
In magnitude measurements of ROTSEIIId,   the magnitude determining accuracy
 decreases for fainter stars (Fig.\ 5 of K{\i}z{\i}lo\u{g}lu et al.\ 2005).
~ \dd ~has a moderate brightness and use of fainter stars than our target
as a reference may significantly contribute to the uncertainties
of individual points in the differential light curve.
When there are suitable stars available within the frame, it is better to choose
 reference stars that are brighter than the target so as not to
 introduce an additional noise into the light curve of the target.
We adopted three stars as the reference whose properties are given
in Table 1, and we used their average light curve in order to reduce long term
systematic errors.
We checked the stability of the reference stars and two test stars.
The mean magnitude of reference stars (see Fig.\ 1) were used in
 the calculation of differential magnitudes.

\subsection{Optical spectroscopic observations}

The spectroscopic observations were performed with RTT150
on May 23, Jun 16, Jul 29, Aug 19 and Sep 26, 2006 using
medium resolution spectrometer TFOSC (T\"UB\.ITAK Faint Object Spectrometer
and Camera). The camera is equipped with a 
 2048$\times $2048, 15$\mu$ pixel Fairchild 447BI CCD.
We used grisms G7 (spectral range 3850-6850 \AA), G8 (5800-8300 \AA), 
G14 (3275-6100 \AA) and G15 (3300-9000 \AA). The average dispersions are 
$\sim$1.5, $\sim$1.1, $\sim$1.4 and $\sim$3 \AA ~pixel$^{-1}$ for G7, G8, G14
 and G15 respectively. We have a low signal to noise ratio for both May and June 2006 
observations due to problems with the autoguider.
The reduction and analysis of spectra were made using 
MIDAS\footnote{http://www.eso.org/projects/esomidas/} and its 
packages: Longslit context and ALICE.
 \begin{figure*}
 \centering
 \includegraphics[clip=true,scale=0.5,angle=270]{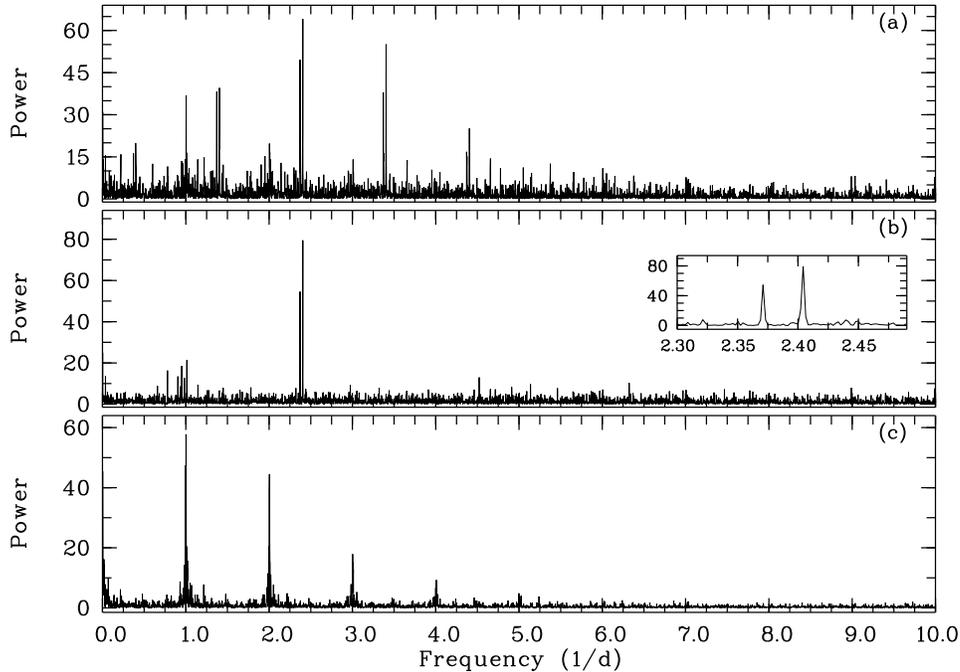}
 \caption{Power spectra for optical counterpart to
          \dd.  Panel (a): Lomb-Scargle algorithm,
          (b): Clean algorithm.
          Frequencies 2.404 and 2.371 \dy are shown in the inset of the
          middle panel for clarity.
          The lower panel (c) is the power spectrum of
          reference stars used in this study.
           }
 \label{fig2}
 \end{figure*}

\section{Long term monitoring}

The differential optical light curve and X-ray light curve of Be/X-ray
binary system \dd ~were shown in Fig.\ 1.
X-ray light curve was obtained from
RXTE/ASM web site\footnote{http://xte.mit.edu}.
 The X-ray data covers the high energy range
(5-12 keV).
Due to a varying amount of circumstellar matter long or short
term variabilities are expected. These variations include both photospheric and
circumstellar variabilities. Therefore a time series analysis for the
observational  data was performed.

\subsection{Variability}

 We searched for a variability in the light curve of Be star by using
several different algorithms:
Lomb-Scargle (Scargle 1982), Clean (Roberts et al.\ 
 1987) and Period04 (Lenz \& Breger 2005)\footnote{available at http://www.univie.ac.at}.
These methods were applied to the whole observational data (from MJD 53499
to MJD 53982).
Frequency analysis was performed
 over a range from 0 to 10 \dy
 ~with a step width 1/T$_{obs}=1/476=0.0021$ \dy. 
Fig.\ 2 shows the power spectra of the Be star (optical counterpart to \dd).
There is a very pronounced window function of one day as expected due to 
a long  observation period. Hence, aliases as a result of window 
function should 
be taken into account.
 \begin{figure}
 \centering
 \includegraphics[clip=true,scale=0.4,angle=0]{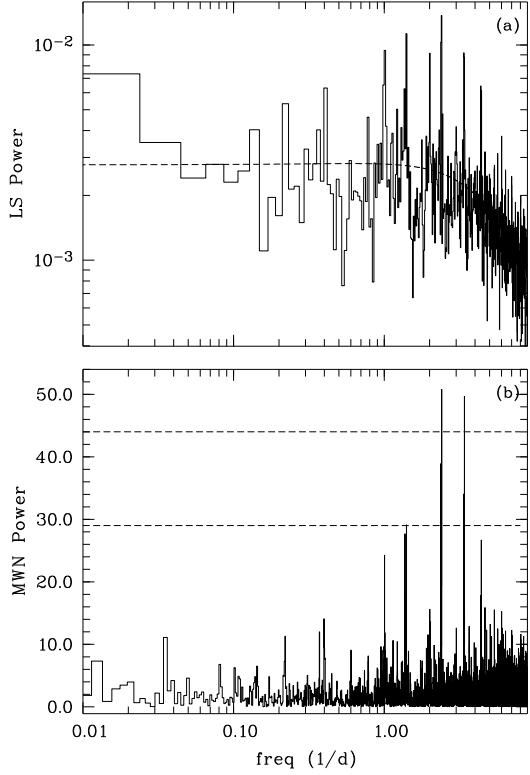}
 \caption{(a): Upper panel is binned LS periodogram of counterpart to
           ROTSEIIId data of \dd.
          A red noise
          component of a very low power index is seen in 1-8\dy ~portion
          of the spectra. Dashed line represents the model continuum
          spectra as explained in the text.
          (b): Below is the same power spectra normalized to the model
          continuum and multiplied by 2 to represent a Poisson distribution
          for 2 degrees of freedom. $3\sigma$ and $5\sigma$ confidence levels
          for the Modified White Noise model are denoted by dashed lines. }
 \label{fig3}
 \end{figure}
 \begin{figure}
 \includegraphics[clip=true,scale=0.30,angle=270]{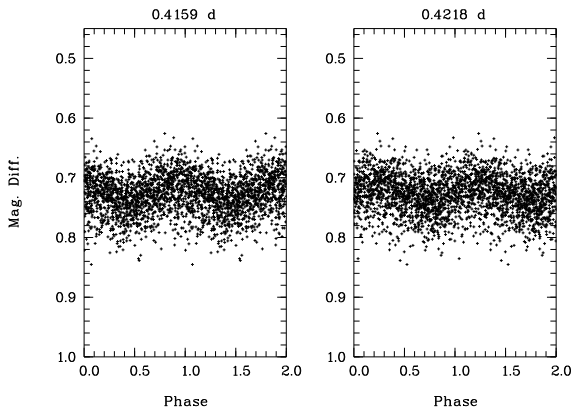}
 \caption{The phase diagrams of 0.416 d and 0.422 d periodic variations
           for the differential light curve shown in Fig.\ 1.   }
 \label{fig4}
 \end{figure}

 As seen from Fig.\ 2, the search of frequencies by
 performing Lomb-Scargle
and Clean algorithms results with the same frequencies. 
The calculated frequencies are 2.404 and 2.371 \dy.
Periodogram analysis with Period04 confirmed these frequencies.
The power spectrum of mean light curve of reference stars obtained by LS 
algorithm is also shown in Fig.\ 2. 
Test stars gave similar results.
Neither the reference stars nor the test stars show pronounced 
periodicities other than one day aliases. There is no sign of
frequencies 2.371 and 2.404 \dy.
However, the significance levels of the oscillations at 2.371 and 2.404  
\dy which are present in both LS and
cleaned spectra, should be set to provide a convincing
evidence of their reality.
 In Fig.\ 3a the log-log binned Lomb-Scargle periodogram of \dd ~is presented.
As one can notice  from the figure, a red noise
component in the power spectra is seen between $\sim$1-8 \dy ~with
a  power law index $f^{-0.004}$. Even though the red noise component has 
a very
low power law index, we modeled the power spectrum in order to test the
significance of these oscillations.
In modeling the broad band continuum spectrum, we rebinned the
power spectrum by a factor of 10.
 Then we modeled the broad band continuum
by a Lorentzian centered at $\sim$ 0.6 \dy ~with a FWHM
$\sim$ 10.04 \dy.  As seen from the figure, in this way, we were able
 to model both low frequency ($< 1.0$  \dy) and
red noise (between 1 and 8 \dy) trends.
 \begin{figure}
 \includegraphics[clip=true,scale=0.3,angle=270]{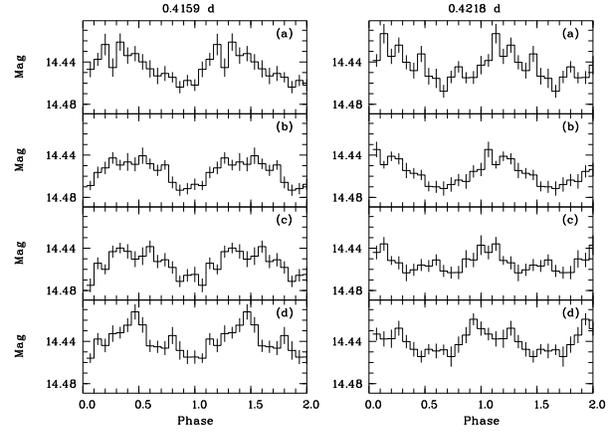}
 \caption{The phase diagrams of 0.416 d and 0.422 d periodic variations
             for the time intervals as mentioned in the text,
             starting from MJD 53499 (a) to MJD 53982 (d). }
 \label{fig5}
 \end{figure}
 \begin{figure}
 \centering
 \includegraphics[clip=true,scale=0.3,angle=270]{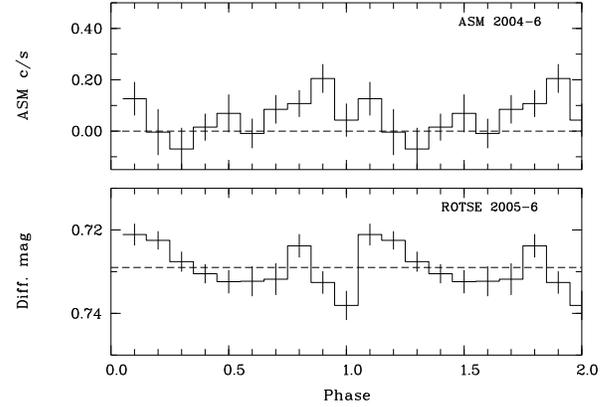}
 \caption{Epoch-folded orbital phase of Be/X-ray binary system
          \dd ~for RXTE/ASM and ROTSEIIId data (using the
          ephemeris $T = $MJD$ 50411.3+55.03N$ given by Wilson et al. 2005). }
 \label{fig6}
 \end{figure}

To test the significance of oscillations, unbinned power spectrum is
divided by its continuum model and the resultant spectrum is
multiplied by a factor of 2 (see van der Klis 1989, and \.Inam et al.\ 
 2004 for applications). The resulting power spectrum
(Modified White Noise model, MWN)
would be a Poisson distribution for 2 degrees of freedom (dof).
 In this presentation, 5\,$\sigma$ detection level of continuum
 normalized power is $\sim$44.
At this level the probability of detecting a false signal is
Q($44|2$)$\sim 2.7 \times  10^{-10 }$.
 Since we have 4822 frequencies (the number of independent 
Fourier steps) in the
power spectra, the probability of having a false signal becomes
$4822 \times  2.7 \times  10^{-10 }$ $\sim 1.3 \times  10^{-6 }$. This
corresponds to the significance of signal detection of
1--$ 1.3 \times  10^{-6 }$ = 0.999999.
In Fig.\ 3b, we present the confidence levels of the observed oscillations.
Only 2.404 \dy has a confidence level greater than 5\,$\sigma$. 
2.371 \dy frequency is in between 3\,$\sigma$ and 
5\,$\sigma$ confidence levels. Other frequencies
are aliasing due to the observation window
and they are successfully removed which can be seen in the cleaned spectra (Fig.\ 2b).

 \begin{table*}
 \centering
 \caption{Journal of spectroscopic observations for H$\alpha$ line.}
 \label{table:2}
 \begin{tabular}{@{}lclcccc@{}}
 \hline\hline
 Date & MJD & Grism$^{\mathrm{a}}$
                & Exp. & FWHM & EW & $\Delta$v (peak sep)\\
      &     &       & (sec) & (\AA) & (\AA) & (km s$^{-1}$)  \\
 \hline
 2006 May 23  & 53878& ~~G7  & 4x600 & 15.8$\pm$6.6 &  2.34$\pm$0.27 & - \\
 2006 June 16 & 53902& ~~G8  & 3x600 & 10.3$\pm$4.3 &  2.42$\pm$0.29 & 539.4 \\
 2006 July 29 & 53945& ~~G7  & 2x1200& 10.0$\pm$4.1 &  2.31$\pm$0.18 & 542.6 \\
              &      & ~~G8  & 2x1800& 11.0$\pm$4.7 &  2.20$\pm$0.17 & 516.9 \\
              &      & ~~G15 & 1800  & 20.1$\pm$8.4 &  2.65$\pm$0.18 & - \\
 2006 August 19& 53966& ~~G8  & 1800  & 10.6$\pm$4.4 &  2.03$\pm$0.19 & 516.1 \\
 2006 September 26& 54004 & ~~G8  & 1800  & 11.1$\pm$4.7 &  2.58$\pm$0.15 &
 465.8\\
              &           & ~~G14$^{\mathrm{b}}$
                 & 3600  &     -        &      - &  - \\
 \hline
 \end{tabular}
 \begin{list}{}{}
 \item[$^{\mathrm{a}}$] Nominal dispersions for G7, G8, G14 and G15 are
           $\sim$1.5, 1.1, 1.4 and 3 \AA pixel$^{-1}$ respectively.
           Hence 3 pixel resolutions at H$\alpha$ line are:
           G7:$\sim$4.5 \AA, G8:$\sim$3.4 \AA, G14:$\sim$4.2 \AA  ~and
           G15:$\sim$9.1

 \item[$^{\mathrm{b}}$] G14 is sensitive in 3275-6100 \AA ~band and
           used for blue spectra.
 \end{list}
 \end{table*}

The phase diagrams of 0.416 d (2.404 \dy) and 0.422 d (2.371 \dy) 
periods calculated from the time series shown in Fig.\ 1 are presented 
in Fig.\ 4. The mean amplitude of the oscillations are 
13.4$\pm$1.2 and 10.4$\pm$1.2 mmag, respectively.
In order to check the stability of the pulsations in 2005 and 2006 we
used target time series so that only the uncertainties of target
measurements will contribute to noise. In Fig.\ 5 we present the same 
phase diagrams calculated from the target time series.
These phase diagrams are obtained by
 dividing the whole data into four parts (part a: MJD 53499-53583,
part b: MJD 53586-53634, part c: MJD 53637-53736, part d: MJD 53836-53982).
Each part has almost the same number of data points in order to get
similar uncertainties in each bin.
The periods are epoch-folded using the epoch T= MJD 53300.
It is shown that the observed periods have coherent phase distributions
 during the different observation spans, in 2005 and 2006.

Short term variations were already found in most of the early type Be stars (Baade 
1982, Rivinius et al.\ 1998, Floquet et al.\ 2002,
Neiner et al.\ 2005). The timescales range from minutes to a few days
and show either the photosphere or the circumstellar environment as
their formation region (Porter $\&$ Rivinius, 2003).
In this study, our calculated frequency of 2.404 \dy ~does not
show  transient character so it can be attributed to the non-radial
pulsation of Be star.
If it was a transient frequency then it should not re-appear with the
same frequency and phase at a later epoch.

No long-term variations were found in the optical light curve.
This shows that disk was not changing very much over the time of 
observations.


\subsection{Orbital period signatures and X-ray quiescence}


Before the year 2002 this system displayed type I outbursts in X-rays
(Wilson et al.\ 2005) due to accretion of matter from Be disk at
periastron passages. Between 2002--2004, type I outbursts
were not seen 
and the orbital period was detected below 3$\sigma$ level
in the work of Wilson et al.\ (2005).
We present 2004--2006 
 RXTE/ASM data (in the energy band 5--12 keV) folded at the orbital period of 
55.03 d using the ephemeris given by Wilson et al.\ (2005)
 ($T = $ MJD $50411.3+55.03N$) in Fig.\ 6. 
In order to see whether there is any orbital period signature
 in optical light curve or not,
we also fold the optical light curve  at a period of 55.03 d 
using the same ephemeris and
we show the result in the bottom panel of Fig.\ 6.
RXTE/ASM data show no correlation with the ROTSE data.
The power spectrum of the ASM data is just noise,
 all of which confirms that the cessation of activity, 
first noted by Wilson et al.\ (2005), has continued in 
subsequent years. 
 \begin{figure*}
 \centering
 \includegraphics[clip=true,scale=0.6,angle=270]{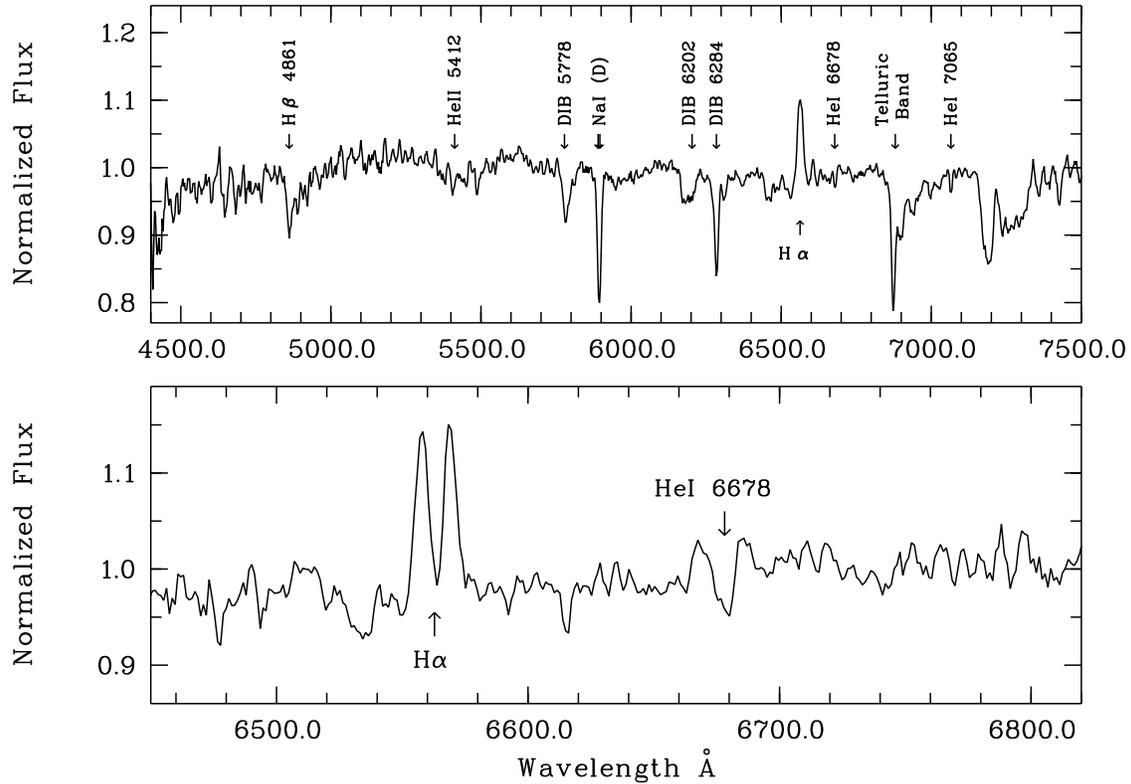}
 \caption{Broadband (4500-7500 \AA) ({\it top}) (with grism  15:
          resolution $\sim9.1$ \AA) and moderate resolution ({\it bottom})
          (with grism 8, resolution $\sim3.4$\AA) spectra of the counterpart
          to \dd ~taken on July 2006.}
 \label{fig7}
 \end{figure*}

\section{Analysis of spectra}

The emission lines of the Be stars particularly H${\alpha}$ ($\lambda$=
6563\AA) can be used for the existence of circumstellar disk of Be stars.
We have observed the counterpart of \dd ~to see whether any
 variation exists on the circumstellar disk or not.
The journal of spectroscopic observations are listed in Table 2.
Broadband (4500-7500 \AA) low resolution spectrum of the
 counterpart is shown in
the upper panel of Fig.\ 7 for July 2006 observation. Several features
are recognized in addition to strong H${\alpha}$ line emission.
In the lower panel of Fig.\ 7, H${\alpha}$ emission with split
profle is visible together with HeI $\lambda$6678 line (the resolution
is $\sim$1.1 \AA ~at H${\alpha}$).
The H${\alpha}$ profiles obtained from each
spectrum are presented in Fig.\ 8. Table 2 gives the measurements of the
equivalent width (EW) of the H${\alpha}$ line for each observing run
together with the values of full width at half maximum (FWHM)
 and the peak separation ($\Delta$v). The FWHM values are obtained from
model calculations of the emission features. The mean FWHM calculated from
G7 and G8 observations of June, July, August and September 2006 
is 10.6$\pm$4.4 \AA.
May and June 2006 observations have relatively low signal to noise ratio.
The H${\alpha}$ line shows a double peaked emission profile
with a mean equivalent width of 2.31$\pm$0.19 \AA ~and
peak separation of 516$\pm$45 km s$^{-1}$.
The mean equivalent width is almost half of the values obtained by
Wilson et al.\ (2005).
H${\alpha}$ line profiles in all runs show a self--absorption at the
center of line due to the
high inclination angle of Be disk. The central absorption with a
varying level in each observation
reaches the continuum on July 2006 observation.
We could not detect whether night to night variations of H${\alpha}$
emission line exist which would imply global structural changes of the disk.

The optical counterpart of \dd ~is in still relative
optic quiescence phase. However the depth of the self absorption feature
in the middle of split H${\alpha}$ emission is deeper than the feature
shown for July 2004 observation of \dd ~by Wilson et al.\ (2005).
This can be interpreted as a decrease in the density of Be disk. 
 Like H${\alpha}$ line, HeI $\lambda\lambda$6678 and 7065 \AA ~lines show
 emission peaks with central absorption.

The blue and mid band spectra in Fig.\ 9
indicate several lines from the Balmer series (H9, H8, H$\epsilon$,
H$\delta$, H$\gamma$, H$\beta$). The typical He I lines,
$\lambda\lambda$4009, 4026, 4121, 4144, 4388 are  shown
 in the lower panel.
Weak or unresolved $\lambda\lambda$4437 and 4472 lines are also 
indicated.
The other HeI lines $\lambda\lambda$4713, 4922, 5016 and 5048 are 
shown in the upper panel of Fig.\ 9.
 Several DIBs are seen in blue, mid and broad band spectra. 

\subsection{Rotational velocity}

Be stars have rapid rotational velocities.
The rotational velocities can be measured by using the empirical
relation derived for FWHM of He I lines (Steele et al.\ 1999).
We calculated the rotational velocity
$v$sin$i$ = 241$\pm$57 km s$^{-1}$
 by using HeI lines $\lambda\lambda$4026, 4143 and 4471.
This velocity is similar to the ones obtained in other
Be systems (Nequeruela $\&$ Okazaki 2001, Reig et al.\ 2004b).
We also calculated the rotational velocity from the average FWHM of
H$\alpha$ lines as $v$sin$i$ = 242$\pm$51 km s$^{-1}$.
However the same velocity derived from the peak separation is
found to be 258 km s$^{-1}$.
On the other hand using the equation given by Hanuschik (1989)
(Eq.1b: empirical relation between FWHM of H${\alpha}$ line and
$v$sin$i$ derived for a sample of 115 Be stars)
we obtained the projected rotational velocity of the H${\alpha}$
emission region  as $v$sin$i$ = 312 km s$^{-1}$.
All these calculations provide similar values.
Calculation of the outer disk radius of the H${\alpha}$
emission region, derived from the peak separation values of H${\alpha}$
line (Coe et al.\ 2006), shows an increase in disk radius during our 
spectroscopic observation period.
 \begin{figure}
 \centering
 \includegraphics[clip=true,scale=0.3,angle=0]{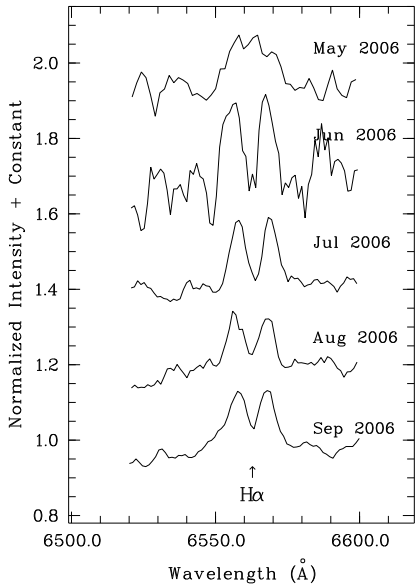}
 \caption{Series of  H${\alpha}$ profiles observed through
             May-September 2006. Note low S/N for May and June data.
             The spectra correspond to the orbital phases of (top to bottom)
             0.01, 0.45, 0.23, 0.61 and 0.30 which are obtained using the
             ephemeris $T$ = MJD $50411.3 + 55.03N$ given by
             Wilson et al. (2005).}
 \label{fig8}
 \end{figure}

 \begin{figure*}
 \includegraphics[clip=true,scale=0.6,angle=270]{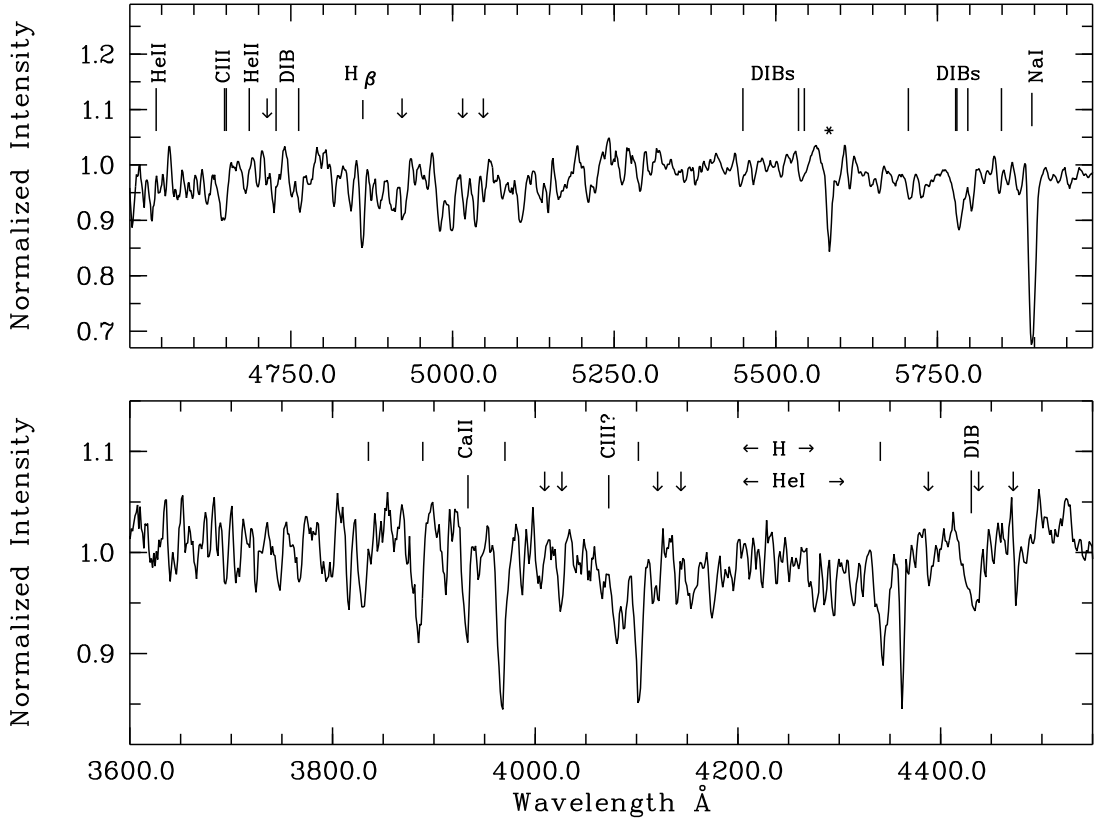}
 \centering
 \caption{ Mid band ({$\it$ top}) and blue ({\it bottom}) spectra
        of optical counterpart to \dd ~taken on July 29
        with grism 7 and on
        September 26 with grism 14.
        Both spectra have been normalized with a spline fit to
        continuum and smoothed with a median filter for display.
        Several strong lines from the Balmer series are indicated by
        short vertical bars. He I lines are shown by down arrows.
        Various other lines and DIBs are also indicated.
        "*" denotes the region affected by a cosmic ray hit.  }
 \label{fig9}
 \end{figure*}

\section{Summary}

Be stars which often display enhancement or fading of their brightness are
candidates for a search of multi-periodicity.
Short term variations has been detected in several Be stars, in most early
type Be stars. Baade (1982) attributed the short term periodic line
 profile variations on timescales between 0.5 and 2 days to non-radial
pulsations. Balona (1990) argued that periods were better explained by
 stellar rotation and he attributed LPV (line profile variability) to stellar spots.
Guerrero et al.\ (2000) reported the optical variability of the Be star
o And during the period 1975-1998 and concluded that neither multimodal
pulsations nor rotational modulation completely explains the complex
light curves. Photometric variations in visual bands on timescales as short 
as one day have been reported by Percy et al.\ (1997) for a sample of Be stars.

Multi-periodicity has been detected in several Be stars in optical LPVs
(Rivinius et al.\ 1998, Floquet et al.\ 2002,
Neiner et al.\ 2005) and in photometric variations 
(Gutierrez-Soto et al.\ 2006) with few hours periods.
 Rivinius et al.\ (2003) reported that the short-term
periodic LPV of Be stars was due to non radial pulsation
which is thought to have a connection with mass loss or circumstellar
disk formation. In addition to this they noted secondary (transient) periods which were attributed
to processes that strongly interact with or reside in the disk.
They were formed in the photosphere and the close circumstellar
environment. These periods were within 10$\%$ of the main photospheric 
period in their sample Be stars.
Multi-periodicity which is mainly detected in optical LPV has been 
generally attributed to non-radial pulsations. Pulsations combined with 
rapid rotation of Be stars are thought to be a prime candidate to explain
mass loss and disk formation in Be stars. 
 In this work, frequency
analysis of the light curve of the Be star counterpart to \dd
~results in one short-term variability. The calculated frequency is
2.404 \dy. This frequency is attributed to non-radial pulsation.
In the study of non radially pulsating Be stars
Rivinius et al.\ (2003) found that photometric periods were detected for stars
which have intermediate to high $v$sin$i$ values. As the amplitude of pulsation
increases $v$sin$i$ value increases for these stars. They noted that
non-radial pulsation induced photometric variability cannot be
seen for the stars whose inclination angle is less (seen pole-on).
In this work the Be star (counterpart to \dd)
has $v$sin$i$ value of about 250 km/s indicating
an inclination angle greater than 40$^{\circ}$ (not to exceed the break-up
velocity of Be stars) and
can be placed well in the defined band of Fig.\ 16 (photometric
amplitude is plotted against $v$sin$i$) of Rivinius et al.\ (2003).

The frequency 2.371 \dy which is very close to 2.404 \dy
as seen in Fig.\ 2 can also be 
thought as a non-radial pulsation of this star. Its confidence is 
$\sim$~4$\sigma$.
These frequencies do not show transient characteristics (Fig.\ 5).
During different observation periods they appear with the same
frequency and the same phase. 
The light curve of the Be star does not show outbursts. \v{S}tefl
 et al.\ (2003)
suggests that transient periods appear during or shortly after 
 outbursts. Therefore pulsations that we found are photospheric.
The absence of outbursts supports the conclusion that the inner
 part of the disk is stable. The disk does not change very much over the
time of observations.

 Periodic variations were not detected in other Be/X-ray binary systems.
 It is believed that the companion has little influence
on the Be star and alters only the outer part of the circumstellar disk.
Then it is possible for Be stars in binary systems to exhibit periodic 
variations. There was a proposed periodic optical variability detection for 
Be/X-ray binary system 2S 0114+65 by Taylor et al.\ (1995) with a period of 
2.77 h and with a semiamplitude of 4 millimag.
The evidence that this
short timescale variation in optical line profiles occurs
 was marginal in the study of Koenigsberger et al.\ (2003). 

We follow up the behavior of H${\alpha}$ line during the five months of 2006
by spectroscopic observations. Double peaked, similar intensity H${\alpha}$
line profiles remain unchanged during this period. We did not observe shell or
single-peaked profiles. The self absorption on emission line, seen in our
spectroscopic observations is caused by
the absorption of light coming from the interior part of the disk, by the
outer regions of the disk. This occurs when the disk is under a large
inclination angle $i$. Be disk is still present.
Our calculated EW values of H${\alpha}$ profiles are smaller than
the calculated values given by Wilson et al.\ (2005)
indicating that Be star disk lost some mass since 2004.
It might also be possible that the density is lower, i.e the disk 
grew and diluted, or a combination of less massive and less dense disk.
 

\begin{acknowledgements}
This project utilizes data obtained by the Robotic Optical Transient Search
Experiment.  ROTSE is a collaboration of Lawrence Livermore National Lab,
Los Alamos National Lab and the University of Michigan
(http://www.rotse.net).
We thank the Turkish National Observatory of T\"UB\.ITAK
for running the optical facilities.
We thank the referee, Th. Rivinius, for a careful reading and valuable 
comments.
Special thanks to Tuncay \"Oz{\i}\c{s}{\i}k and colleagues from TUG
who keeps hands on ROTSEIIId.
We acknowledge support from T\"UB\.ITAK, The Scientific and Technological 
Research Council
of Turkey,  through project 106T040.
We also acknowledge the RXTE/ASM team for the X-ray monitoring data.
\end{acknowledgements}

\end{document}